\documentstyle[aps,preprint]{revtex}
\begin{document}
\draft
\preprint{IMSc-96/16; hep-th/9608170}
\title{Fermions in Black Hole Space-Time: Hawking Radiation and Back
Reaction}
\author{Arundhati Dasgupta and Parthasarathi Majumdar
\footnote{E-Mail:dasgupta,partha@imsc.ernet.in}}
\address{The Institute of Mathematical Sciences, \\ CIT Campus,
Madras - 600 113,  India.}
\maketitle
\begin{abstract}
The leading order correction to the metric of a Schwarzschild 
black hole, due to the backreaction of infalling fermionic matter fields,  
is shown to produce a shift of the event horizon such that particles 
that would constitute Hawking radiation at late retarded times are now 
trapped. Fermionic field operators associated with 
infalling and outgoing modes at the horizon behave canonically in 
the semiclassical approximation. They are, however, shown to satisfy a 
nontrivial exchange algebra given in terms of the backreaction, 
when the shift is `quantized' by means of correspondence. The consequent 
exchange algebra for bilinear fermionic densities is also obtained. 

\end{abstract}

\section{INTRODUCTION}
\label{sec:int3}

It is well-known that fermionic matter behaves quite differently from scalar quantum fields 
in black hole spacetimes. The foremost difference stems from the 
fact that fermionic fields are not subject to the weak energy condition 
\cite{hawlis}
$T_{\mu \nu} \xi^{\mu} \xi^{\nu} > 0$ on the energy-momentum tensor 
$T_{\mu 
\nu}$, where $\xi^{\lambda}$ are time-like Killing vectors. An important 
consequence of this difference is the absence of super-radiant scattering of
fermions off charged or rotating black holes \cite{unr}. The spin
statistics theorem which determines the property of the energy momentum tensor
stated earlier, implies that there
can be only one particle per each energy state of the fermion. It follows 
that the amplitude of the reflected wave cannot be enhanced by the black 
hole, thereby preventing
the black hole from losing angular momentum or charge by fermion 
emission. The dissimilarity crops up again in the thermal radiation 
from evaporating  black holes \cite{hawk}, where the spectral distribution 
from such a black hole, 
upon identifying its surface gravity with the Hawking temperature, is a 
Fermi-Dirac distribution for massless fermionic fields rather than the 
Planckian distribution expected for scalars. 

An aspect of Hawking's seminal analysis of black hole radiation that has 
since received attention from diverse angles is the issue of backreaction 
of matter fields on the black hole geometry. In the original (asymptotic) 
analysis, this was considered only in an adiabatic approximation whereby 
the thermal radiation caused a reduction in the mass of the radiating 
black hole resulting in an increase of the Hawking temperature (and hence 
in the intensity of radiation). The consequent shrinking in the horizon 
size eventually leads to a {\it complete} evaporation of the black 
hole, leaving behind only thermal radiation, -- a process which, if true, 
implies a breakdown of standard quantum mechanics with its law 
of unitary evolution. This is the notorious Information Loss Paradox. 
The resolution of the paradox may in principle entail modification of the 
principles of quantum mechanics \cite{haw2}, or indeed consideration of 
{\it remnant} matter that retains the information (in terms of pure quantum 
states) that fell into the black hole \cite{gidd}, \cite{gidd1}. A less exotic 
approach, one which endorses standard quantum mechanics and seeks to 
discern aspects of the original analysis that may be subject to 
modification upon inclusion of effects neglected earlier, may be another engaging 
possibility \cite{hoof1} - \cite{verl}. This more conservative viewpoint will be adopted 
in what follows. 

According to this premise,
quantum states characterizing the horizon differ sharply from 
the benign vacuum configuration corresponding to a fixed classical 
background considered in the purely semiclassical 
approach. The horizon is more likely a dense gas of strongly interacting 
particles at very high energies \cite{hoof1}, \cite{englr} -- possibly even a gas of open 
strings \cite{suss}. Infalling particles interact with this system 
according to the laws of unitary evolution, leading to a 
(possible) description in terms of a quantum mechanical S-matrix 
\cite{hoof1} involving only gravitational interactions. The large 
blue-shifts, close to the horizon, on the solutions of the matter 
field equations in the black hole geometry (which incidentally is germane 
to Hawking's asymptotic analysis \cite{hawk}) stand to interpretation in 
terms of the corresponding field quanta acquiring enormous centre-of mass 
energies. Momentum transfers, on the other hand, are taken to be low, on 
account of the large angular separation expected between the most 
frequently colliding particles. Thus, gravitational interactions tend to 
dominate over all others in these kinematics.\footnote{Admittedly, such 
kinematical arguments carry a hint of flat space intuition.}

These interactions have been considered for scalar matter fields by Kiem 
et. al. \cite{verl}. The authors consider linear perturbations of the 
metric due to matter backreacting classically on the geometry. 
The modified geometry turns out to be the `cut and paste' 
of two Schwarzschild geometries along the horizon (given by the 
Kruskal coordinate $x^+=0$) 
with a shift $\delta x^-$ which is given in terms of the classical 
backreaction \cite{dray}. This shift, in turn, can be transcribed into a 
shift of the reference null Eddington-Finkelstein (EF) coordinate 
(defining the optimum `time' at which a massless particle must leave ${\cal I}^-$ so 
as to graze the horizon and just make it to ${\cal I}^+$), thereby 
affecting outgoing Hawking particles: the sign of the shift is such that 
these particles are now trapped. Consequently, the infalling and outgoing 
modes of the matter field operator at the horizon, which, in the absence of 
backreaction are 
independently measurable observables, are now no longer so. Instead, their 
commutator is proportional to the shift of the reference null EF 
coordinate. A more nontrivial exchange algebra is derived between these 
modes, by appealing to a correspondence principle to go beyond the 
semiclassical approximation, by promoting the shift to an operator 
(albeit a bilocal one).\footnote{Such an algebra between two observables 
which are 
usually taken to be mutually commuting is strongly reminiscent of  
Complementarity in quantum mechanics. Hence the name {\it Black Hole 
Complementarity} for this result.}

In this paper, we address the issue of backreaction of neutral massless {\it 
fermionic} 
matter falling on the black hole geometry. In particular, we focus on the 
anticommutator of the 
infalling and outgoing modes of the fermionic field operator at the 
horizon of a Schwarzschild black hole, to determine whether these modes 
interact gravitationally just like scalar matter. We show that the 
horizon undergoes a shift quite similar to the scalar case, in a 
linearized Einstein-Cartan formulation. Consistent propagation of 
fermions in the modified geometry leads to a {\it canonical} anticommutation 
relation between these modes in the semiclassical approximation. This leads to a canonical 
algebra for fermionic bilinear densities. Appealing next to the 
correspondence principle, a non-trivial 
anticommutation exchange algebra is derived for the fermionic modes. Fermionic 
bilinear densities satisfy an exchange algebra similar to the one found for scalar matter.

The paper is organized as follows:  In the second section we briefly review 
the Hawking process for
fermions. In section 3 we use a perturbed metric to model the effect of
the back reaction of the fermion energy momentum tensor. In section 4
the effect of this is studied and the non-trivial effect on promoting
the perturbation as a operator-valued quantum fluctuation is investigated. Finally in the 
concluding section we discuss the limitations of the method used, 
and the issues yet to be addressed in this approach.

\section{Weyl Fermions in a Schwarzschild black hole}

We begin with the Weyl equations for massless fermions in a Schwarzschild 
spacetime; the metric of the latter is given, in Kruskal-Szekeres 
coordinates, by  
\begin{equation}
ds^{2} = {32(GM)^{3}\over {r}}{e^{-{r}/{2GM}}}  dx^{+} dx^{-} - 
r^{2}d\Omega^{2}~. \label{sch}
\end{equation}
A choice of tetrad components $e^{m~ \alpha \dot{\beta}}$ reproducing 
this metric is given by \cite{unr}
\begin{eqnarray} 
e^{x^{+} \dot{+} +}=& -{x^{+}\over{2GM \Delta^{1/2}}}&\\
e^{x^{-}\dot{-} -}=&{x^{-}\over{2GM\Delta^{1/2}}}&\\
e^{\theta~ \dot{-}+}=&- 1/r& ~~~~e^{\theta~ \dot{+}-}= -1/r\\
e^{\phi~ \dot{-} +}=& -\imath /{r \sin{\theta}}& ~~~~e^{\phi
~\dot{+} -}=\imath/{r \sin{\theta}}~.\label{tetr}
\end{eqnarray}
Here, $\Delta~\equiv~(1 - {2GM \over r})$. The dotted-undotted pair 
of indices indicate a tangent space vector as 
usual, with dotted (undotted) indices {\it per se} indicating chiral 
(antichiral) spinors of the tangent space Lorentz group. The Weyl 
equation is then given by,
\begin{equation}
\imath \nabla^{\alpha {\dot \beta}} \psi_{\alpha}~=0~=~ 
\imath \nabla^{\alpha {\dot \beta}} {\bar \psi}_{\dot \beta}~, \label{weyl}
\end{equation}
where 
\begin{equation}
\imath \nabla^{\alpha {\dot \beta}} \psi_{\alpha}~\equiv~\imath e^{m, 
\alpha \dot{\beta}} \left(~\partial_{m} 
\psi_{\alpha} -\omega_{m \alpha}^{~~\gamma}~\psi_{\gamma}\right)~etc.
\end{equation}
and $\omega_{m \alpha}^{\beta}$ are the (chiral) spin connection matrices, 
given in terms of the tetrad components by the standard formula
\begin{equation}
\omega_{m \alpha}^{\gamma}~\equiv~e^n_{\alpha {\dot \beta}} e^{\gamma 
{\dot \beta}}_{n ; m}~.
\end{equation}
Near the horizon, $\Delta \rightarrow 0$ ; keeping in 
mind that length (and time) as measured by asymptotic observers scale by 
the singular factor $\Delta$ in the horizon region, because of infinitely 
large blue-shifts that the solutions of the dynamical equations undergo, one 
can rescale $\psi \rightarrow \psi/\Delta^{\frac12}$. This reduces the pair 
(\ref{weyl}) to
\begin{eqnarray}
\partial_v \psi_+~&=&~0~=~\partial_v {\bar \psi}_{\dot +} \label{lfm} \\
\partial_u \psi_-~&=&~0~=~\partial_u {\bar \psi}_{\dot -}, \label{rtm} 
\end{eqnarray}
where $u, v$ are the Eddington-Finkelstein null coordinates. 
Thus, the Weyl field decomposes into `retarded' (outgoing) and `advanced' 
(incoming) solutions near the horizon, similar to the scalar field \cite{verl}
\begin{eqnarray}
\psi^{out}~&=&~\psi_+(u, \Omega)~\label{out} \\
\psi^{in}~&=&~\psi_-(v, \Omega) ~.\label{in} 
\end{eqnarray}
Since the {\it out (in)} solutions constitute a complete set of solutions 
to the Weyl equation on ${\cal I}^+~~({\cal I}^-)$, one can match the 
advanced and retarded propagation of these solutions at the horizon a la' 
Hawking \cite{hawk}, leading to the reparametrization
\begin{equation}
\psi^{out}(u(v), \Omega)~~=~~\psi^{in}(v, \Omega)~, \label{reprm} 
\end{equation}
where, $u(v)=v_0-4GM \log({{v_0 -v} \over 4GM})$. The  
reparametrization is singular at $v_{0}$ which represents
the latest reference 'time' at which an incoming wave leaves ${\cal 
I}^-$  to get scattered to 
$\cal I^{+}$ along the event horizon given by $u~\rightarrow~\infty$. 
{}For $v > v_0$, all incoming waves are trapped by the black hole. 

Following the original treatment \cite{hawk}, the reparametrization above 
is next used to compute the (asymptotic) Bogoliubov coefficients 
appearing in the Bogoliubov transformations connecting the 
creation-annihilation operators of fields having support on the two 
asymptotic null infinities
\begin{equation}
c_{\omega}~~=~~\int d \omega'~(~\alpha_{\omega \omega'} 
a_{\omega'}~+~\beta_{\omega \omega'} b^{\dag}_{-\omega'}~)~,\label{bog}
\end{equation}
where $c~,~c^{\dag}~(a~,~a^{\dag}~)$ are the creation-annihilation 
operators associated with $\psi_{out}~(~\psi_{in})$ on ${\cal I}^+~({\cal 
I}^-)$. The distinction from the scalar case now manifests as one 
attempts to calculate the spectral distribution of the outgoing radiation 
by calculating the expectation value of the number operator $c^{\dag} c$ 
in the vacuum on ${\cal I}^-$; here one must recall that these operators 
obey an anticommutation algebra instead of a commutation relation. An 
immediate consequence is a thermal spectrum with a Fermi-Dirac 
distribution, upon identifying the surface gravity with the Hawking 
temperature \cite{hawk}, \cite{unr}. 

This analysis has obviously ignored the change in the black hole geometry 
induced by infalling and outgoing fermionic matter -- the backreaction. 
In the next section we attempt to incorporate it in the semiclassical 
approximation: the linearized change in the black hole tetrad components 
due to infalling fermionic matter is determined, and re-expressed as a 
shift in the horizon. This is then used to determine an exchange algebra 
for the fermionic fields. 

\section{Classical Backreaction For Fermions}
The dominant effect of backreaction {\cite{arn}}of quantum fermionic matter on the classical black 
hole geometry can be characterized by a linearized perturbation of the frame 
components: $e_m^{\alpha 
{\dot \beta}} \rightarrow e_m^{\alpha {\dot \beta}} + h_m ^{\alpha {\dot \beta}}$. The 
linearized fluctuations $h_m^{\alpha {\dot \beta}}$ are related to linearized 
fluctuations of the Schwarzschild metric according to 
\begin{equation}
h_{mn}~=~ e_{(m}^{\alpha {\dot \beta}} h_{n), \alpha {\dot \beta}}~,
\end{equation}
where the $e_{m}^{\alpha {\dot \beta}}$ are the Schwarzschild tetrad components given 
in (\ref{tetr}). We are especially interested in the effect of infalling fermionic 
fields on the black hole geometry. Following Hawking's approach to black hole 
radiation \cite{hawk}, it is sufficient to restrict to waves of very high frequency near 
the horizon, i.e., adopt the geometrical optics approximation. For fermion fields 
falling on the horizon, therefore, the largest contribution to the backreaction will 
come from $T_{x^+, + {\dot +}}$ and $T_{x^-, - {\dot-}}$, the
nonzero components of the energy momentum tensor in the
longitudinal direction, where $T_m^{\alpha {\dot \beta}}$ may be 
taken to be (the expectation value of) the fermionic energy momentum tensor near 
the horizon. The other energy-momentum density components are negligible in the 
kinematical regime of interest.

Appealing now to the Einstein-Cartan equation
\begin{equation}
R_{m}^{\alpha {\dot \beta}}~-~\frac12 e_m^{\alpha {\dot \beta}} R~=~T_m^{\alpha {\dot 
\beta}}~~, \label{eceq} \end{equation}
the tetrad fluctuations are seen to obey the linearized Einstein-Cartan equation (see 
Appendix A)
\begin{equation}
(~\nabla_{\Omega}^2~-~1~)h_{x^+,+ {\dot +}}~=~k~T_{x^+, + {\dot +}}~~, \label{eq:lec}
\end{equation}
($k$ is a constant depending upon the mass of the black
hole), with the solution 
\begin{equation}
h_{x^+, + {\dot +}}~=~\int d^2 \Omega'~f(\Omega~,~\Omega')~T_{x^+, + {\dot +}} ~,
\label{pert}
\end{equation}
where, $f(\Omega, \Omega')$ is the Green's function of the Laplacian on the two-sphere 
\cite{dray}. With this modification of the geometry near the black hole horizon, 
consistent propagation of Weyl fermion fields, very close to the horizon, is described 
by the modified equations (cf. eq.s (\ref{lfm}, \ref{rtm}))
\begin{equation}
x^- \nabla_{x^-} \psi_-~=~0~=~x^+ \nabla_{x^+} \psi_+~~, \label{modeq}
\end{equation}
where, $\nabla_{x^+} \equiv \partial_{x^+} - h_{x^+x^+} \partial_{x^-}$ etc. Where $h_{x^+ x^+}= e^{+ \dot{+}}_{x^+}~h_{x^+,+
\dot{+}}$. The 
formal solution for $\psi^{out} \equiv \psi_+$ may be written as
\begin{equation}
\psi_+~=~\psi_+(x^-~+~\int^{\infty}_{x^+_0}~ dy^+ h_{x^+x^+}(y^+, \Omega))~ .
\label{qout}
\end{equation}
A similar solution exists for $\psi^{in} \equiv \psi_-$ with a shift in the other Kruskal 
coordinate $x^+$. The important point to note in (\ref{qout}) above is that the 
integration limit excludes the interval $(0, x^+_0)$; as pointed out in \cite{verl}, this 
region is not interesting for our purpose of estimating the effects of backreaction, 
since for $v < v_0$, all infalling waves reflect back onto ${\cal I}^+$. Backreaction 
effects are important only for particles that get trapped behind the horizon. 

Relating the shift $\delta x^-$ given in (\ref{qout}) to the affine parameter $\lambda$ of 
the null geodesic generator of the {\it modified} event horizon (obtained to linear order 
in the shift of the tetrad components), one can obtain a la' Hawking \cite{hawk} a 
matching condition for $v > v_0$ between the incoming and outgoing solutions
\begin{equation}
\psi^{in}(v(u)+\delta v_0)~=~\psi^{out}(u)~, \label{mach}
\end{equation}
where,
\begin{eqnarray}
v(u) &= & v_{0} - 4GM e^{(v_{0}-u)/4GM} \\
{}&{} & {} \nonumber \\
{ \delta v_{0}} & = & 
{-}{\int_{v_{0}}^{\infty}dv}\int{d\Omega~f(\Omega,\Omega')} {e^{(v_{0}-v)/4GM}}T_{v 
v}\;\;\; \mbox{$v> v_{0}$} \label{eq:shift}
\end{eqnarray}
Thus, the shift in the optimum value of the EF coordinate on ${\cal I}^-$ is such that 
fermions that would have made it to ${\cal I}^+$ in the absence of backreaction, are now 
trapped behind the shifted horizon. A remark on the similarity of the shift 
$\delta v_0$ found here and that found in \cite{verl} for scalar matter is perhaps in 
order. Recall that near the horizon infalling particles are blue-shifted to enormously 
high energies, so that one can appeal 
to the geometrical optics approximation in dealing with these. Now in flat spacetime, it 
has been shown \cite{maj} that in this approximation (which corresponds to the eikonal 
approximation), fermionic and scalar cross sections become identical for electromagnetic 
interactions, pointing to an on-shell induced supersymmetry. The similarity between the 
gravitational effect of fermions and scalars on the horizon seen here may be attributed to 
such a supersymmetry. Clearly, in the geometrical optics approximation, helicity-flip 
amplitudes for Weyl fermions vanish. 

\section{Effects of Backreation}
\subsection{Semiclassical Approximation}

The semiclassical approximation, as always, considers quantum (fermionic) matter in a 
classical gravitational background, which in this case is a (backreaction-modified)
Schwarzschild geometry. In the absence of backreaction, scalar field operators 
corresponding to 
{\it out} and {\it in} solutions of the covariant Klein-Gordon equation are known to be 
mutually commuting for $v > v_0$ \cite{verl}. It is easy to see that, with the modification 
discussed above to the geometry, this is no longer the case: to linear order in the 
(c-number) shift $\delta v_0$, one can show that
\begin{equation}
~[\phi^{out}(u(v), \Omega)~,~\phi^{in}(v', \Omega')]=2 \pi i\delta 
v_{0}~\delta(v-v')~\delta^{(2)}(\Omega - \Omega')~. \label{sclr}
\end{equation}
Here use has been made of the canonical commutation relation for scalar fields \cite{verl}
\begin{equation}
[\phi^{in}(v_1, \Omega_1)~,~\partial_{v_2} \phi^{in}(v_2, \Omega_2)]~=~2\pi i 
\delta(v_{12})~\delta^{(2)}(\Omega_1 - \Omega_2)~. \end{equation}

A similar analysis for fermionic field operators, using the canonical {\it 
anti}commuting relations
\begin{eqnarray}
\{~\psi_{\alpha}(v_1, \Omega_1)~,~\psi_{\beta}(v_2, \Omega_2)~\}~&=&~0~,~ \alpha, \beta=+, 
-~\nonumber \\ 
\{~{\bar \psi}_{\dot \mp}(v_1, \Omega_1)~,~ \psi_{\mp}(v_2, \Omega_2)~\}~&=&~2\pi i 
\delta(v_{12})~\delta^{(2)}(\Omega_1-\Omega_2)~\label{car}
\end{eqnarray}
leads to a null result to linear order in $\delta v_0$:
\begin{equation}
\{~\psi^{out}~,~\psi^{in}~\}~=~0 ~\label{scl}
\end{equation}
as one might have expected. Indeed, fermionic fields are themselves unobservable; 
densities constructed out of fermionic bilinears are the true observable quantities. 
The simplest  
bilinear composites constructed out of fermions are the
components of the current in the tangent plane $J_{\alpha \dot {\beta}}$. The dominant  
components of the current are\footnote{The other components are 
negligible in the kinematical situation under consideration. Also, scalar bilinears 
vanish for chiral fermions.}
\begin{equation}
J^{in}(v)~\equiv~J_{-{\dot -}}~\equiv~\psi_-(v) {\bar \psi}_{\dot -}(v) 
~,~ J^{out}(u)~\equiv~\psi_+(u) {\bar \psi}_{\dot +}(u) ~, \end{equation}
These bilinear observables turn out also to behave canonically, i.e.,  $[J^{out}, 
J^{in}] = 0 $ to lowest order in $\delta v_0$ (see Appendix B for a sketch of the 
commutator calculation). This behaviour is quite in contrast to the behaviour of scalar 
fields in the semiclassical approximation: the infalling and outgoing modes of 
the fermion field appear to be completely independent, i.e., non-interacting, despite 
the classical shift of the horizon. 

\subsection{Quantum Correspondence and Exchange Algebras}

With fields as operators, the energy-momentum flux operator\footnote{The Energy Momentum  
tensor is not well-defined near the horizon, but the momentum flux is \cite{hawk}.} 
$P_{v} \equiv \int_{v_{0}}^{\infty}~dv~T_{v v}$ has a
nontrivial commutation relation with the incoming field. It generates
translations along the $v$ direction. 
\begin{equation}
[P_{v}(v),\psi^{in}(v')]= 2\pi\imath \delta^2{(\Omega
-\Omega')}\partial_{v}\psi^{in}(v',\Omega')
\end{equation}
Now, as seen in equation (\ref{eq:shift}), the shift $\delta v_{0}$ is related to the energy
momentum tensor. If we assume that this classical relation can be
promoted as such to a relation between operators by means of some sort of a quantum 
correspondence principle \cite{verl}, it can be used
find the commutation relation of $\delta v_{0}$ with the incoming field
\begin{equation}
[\delta v_{0}(\Omega)~,~\psi^{in}(\Omega')] =
-16\pi\imath f(\Omega,\Omega')e^{(v_{0}-v)/4GM}\partial_{v}\psi^{in}\;\;
\;\;\;
\mbox{$v>v_{0}$}~.
\label{eq:comm}
\end{equation}
This result can now be used, following Kiem et. al. \cite{verl} to determine an exchange 
algebra between the {\it in} and {\it out} field operators. Keeping in mind the canonical 
anticommutation relations obeyed by the fermion fields, this algebra (to lowest order in 
the backreaction) is given by
\begin{equation}
\psi^{out}(u, \Omega)\psi^{in}(v, \Omega')=
-\{~1~-~16\pi\imath 
f(\Omega,\Omega')e^{(u-v)/4GM}\partial_{u}\partial_{v}~\})\psi^{in}(v, 
\Omega')\psi^{out}(u,\Omega)~. \label{eq:exc}
\end{equation}
In the absence of backreaction $f(\Omega, \Omega')=0$, we get the standard anticommutation 
relation between the field operators. 

Thus to the extent one can trust the procedure of promoting (\ref{eq:shift}) to the level 
of an operator relation, fermionic field operators appear to obey the requirements of 
Complementarity. The point becomes clearer when one computes, within the same 
correspondence approach, the exchange algebra of fermionic bilinear densities; this turns 
out to be similar to the exchange algebra of scalar fields \cite{verl}
\begin{equation}
J^{out}(u, \Omega)J^{in}(v, \Omega')=
\{~1~-~16\pi\imath  
f(\Omega,\Omega')e^{(u-v)/4GM}\partial_{u}\partial_{v}~\} J^{in}(v, \Omega')J^{out}(u, 
\Omega) ~. \end{equation}
In other words, observables, mutually commuting in absence of backreaction, now obey a 
nontrivial exchange algebra (for $v > v_0$). The similarity of the exchange algebras 
for the fermion bilinears and scalar fields is 
of course, no surprise, being expected on account of the similarity in the behaviour of 
fermions and scalars 
under spacetime translations which lies at the root of the derivation of these algebras. 
In addition, as already mentioned, since the geometrical optics approximation is 
essentially being used here, and fermions and scalars behave similarly in this 
approximation, the only difference that may be expected is from the statistics, i.e., the 
fact that fermionic field operators obey a canonical anticommutator algebra. This is as 
obvious here as in the case of the spectral distribution of Hawking radiation.

\section{Conclusions}

In the semiclassical approximation, backreaction of matter does not seem to induce any 
interaction between infalling and outgoing modes of the fermion field, while for scalars 
such interactions do indeed appear. It is important to keep in mind however that in 
this analysis, tranverse gravitational 
interactions as well as non-gravitational forces have been ignored on the plea that they 
would be subdominant in this situation. Such forces could conceivably change this 
results somewhat \cite{englr}, although a complete picture can 
only be expected after nonperturbative `quantum gravity' is understood. In any case, our 
analysis does not indicate any serious flaw with the semiclassical approximation 
in the fermionic case, even though for scalars going beyond this approximation seems 
imperative. A deeper understanding of this difference, possibly in terms of foliations 
of the geometry (e.g., along lines suggested in \cite{mathr}) is certainly desirable. It 
might be related to the diffrence mentioned at the outset, namely that fermions do not 
obey the weak energy condition. 

As mentioned earlier, the validity of the exchange algebras derived above hinges on the 
assumption that the shift $\delta v_0$, while a 
shift in the optimum value of a coordinate, can be elevated to the level of an operator on 
Fock space. Clearly, as an operator, $\delta v_0$ is bilocal which might possibly 
underlie a justifiable suspicion of 
a violation of microcausality. In the scalar field case, Kiem et. al. introduce a third 
scalar field $\phi^{hor}$ to ensure that microcausality is maintained, with $\phi^{in}$ 
evolving unitarily to it at the horizon from ${\cal I}^-$. It is not clear whether this is a 
valid procedure without a full quantum theory of gravity to back it up. We therefore adopt a 
deliberately ambivalent stand on this issue. 

In fact, even promoting equation (\ref{eq:shift}) to the level of an operator equation
and thereby endowing operatorial status to a coordinate, purely by appealing to a
correspondence principle, is questionable in a situation where a well-defined quantum
theory is yet unknown. Usually, a correspondence principle works best in the reverse
manner, i.e., in discerning the structure of a classical theory {\it corresponding} to a
known quantum theory. The application of correspondence in the present context (which
elevates the linearized Einstein equation to the level of an operator equation), while
going beyond the purely semiclassical approximation (where gravity is classical), appears
to be replete with ambiguities, again in the absence of a proper `quantum geometry'. On
the other hand, the super-Planckian energies acquired by field quanta near the horizon
does seem to warrant an approach beyond the semiclassical approximation. In this context,
a better approach than the one adopted here might be to follow the quantization of
monopoles in gauge field theory, using the method of collective coordinates. Indeed, the
horizon, as already mentioned, is modeled by a dense two dimensional gas of strongly 
interacting particles. Such a system, like a large nucleus, is very likely to
undergo collective motion which should be amenable to exact analysis, unlike the
(linearized) metric (or tetrad) fluctuations dealt with perturbatively here (the latter
are analogous to Gaussian fluctuations around the ground state in the case of the
nucleus). We hope to report on these issues elsewhere.

\acknowledgments

We thank S. Das, S. R. Das, P. 
Ramadevi and T. Sarkar for helpful discussions during the course of this work. A. D. 
would like to thank G. Date for bringing ref. \cite{unr} to her attention. P. M. would 
also like to thank the Saha Institute of Nuclear Physics for the kind hospitality 
extended to him during a visit during which this work was being completed, and A. P. 
Balachandran, A. Chatterjee, A. Ghosh, K. S. Gupta, P. Mitra and V. V. Sreedhar for 
illuminating discussions.

\appendix\section{}

The linear order perturbation to the tetrads is shown to
satisfy equation (\ref {eq:lec}) here.
The tetrad is perturbed to linear order $e^{\alpha
\dot{\beta}}_{m}\rightarrow e^{\alpha \dot {\beta}}_{m}+ 
h^{\alpha \dot{\beta}}_{m}$. Hence, to linear order in $h$, the
two form,
\begin{eqnarray}
R_{~m n}^{\alpha \dot{\beta},\gamma \dot{\delta}}&=&\partial_{m}~
\delta\omega_{n}^{\alpha \dot{\beta},\gamma \dot{\delta}} -~
\partial_{n}~\delta\omega_{m}^{\alpha \dot{\beta},\gamma \dot{\delta}}
~+~ \delta\omega_{m}^{\alpha \dot{\beta},~ \alpha' \dot{\beta'}}
 \omega_{n,\alpha' \dot{\beta'}}^{\gamma \dot{\delta}}~+~
\omega_{m}^{\alpha \dot{\beta},~ \alpha' \dot{\beta'}}
 \delta\omega_{n,\alpha' \dot{\beta'}}^{\gamma \dot{\delta}}\nonumber \\
&& -~\delta\omega_{m,\alpha' \dot{\beta'}}^{\gamma \dot{\delta}}
 ~\omega_{n}^{\alpha \dot{\beta}, \alpha' \dot{\beta'}}
~-~\omega_{m,\alpha' \dot{\beta'}}^{\gamma \dot{\delta}}
~\delta \omega_{n}^{\alpha \dot{\beta}, \alpha' \dot{\beta'}}~.
\label{change}
\end{eqnarray}
Where $\delta \omega$ stands for the linearised spin
connection. The change in the tetrad, is only along the
longitudinal direction as near the horizon the energy momentum
tensor of the matter field is dominant in these directions
only.
\begin{equation}
\delta e_{x^+,+ \dot{+}}= h_{x^+,+ \dot{+}}(x^+,x^-,\Omega),~~~ \delta e_{x^-,- \dot{-}}=h_{x^,- \dot{-}}(x^+,x^-,\Omega) 
\end{equation}
The changes in the spin connection very near the
horizon (neglecting terms proportional to $x^+$ and higher
powers),are calculated and the relevant self dual parts are:
\begin{eqnarray}
(\delta \omega_{x^+})_{\pm}^{~\pm}&=&\mp{h_{x^+, + \dot{+}}\over
{\Delta^{1/2}}}\\
(\delta \omega_{x^+})^{~-}_{+}&=&-{1\over r} \left(\partial_{\theta}
+ {\imath \over
{\sin{\theta}}}\partial_{\phi}\right)~h_{x^+,+ \dot{+}}
\end{eqnarray}
Similar equations are obtained for $\delta \omega_{x^-}$.
The spin connections of the original metric were,
\begin{equation}
(\omega_{x^+})_{ \pm}^{~\pm}= \pm{1\over
{2x^+}}{4(GM)^{2}\over {r^2}},\;\;\;\;\;
(\omega_{x^-})_{\pm}^{~\pm} = \pm{1\over
{2x^-}}{4(GM)^{2}\over {r^2}}
\end{equation}
\begin{equation}
\omega_{\theta,\pm}^{~~~\mp}= \mp~\Delta^{1/2},\;\;
\omega_{\phi,\pm}^{~~~\pm}= \mp~ \imath \cos{\theta},\;\; 
\omega_{\phi,\pm}^{~~~\mp}= \imath \sin{\theta}\Delta^{1/2}~.
\end{equation}
Using the above the perturbed curvature can be calculated
from equation {\ref{change}}. The contraction with the
vierbein gives $R_{m,\gamma \dot{\delta}}=e^{n, \alpha
\dot{\beta}}R_{m n,~\alpha \dot{\beta} \gamma
\dot{\delta}}$. From Einstein-Cartan equation (\ref{eceq}), equation (\ref{eq:lec}) 
follows.
\section{}
The commutator of the currents are calculated to order $\delta
v_{0}$.
\begin{eqnarray}
~\left[J^{out},J^{in}\right]&=&\left[\bar{\psi_{\dot{+}}}\psi_{+},\bar{\psi_{ 
\dot{-}}}\psi_{-}\right]\nonumber \\
&=&\bar{\psi_{\dot{+}}}\left\{\psi_{+},\bar{\psi}_{\dot{-}}\right\}\psi_{-}
-\left\{\bar{\psi_{\dot +}},\bar{\psi_{\dot{-}}}\right\}\psi_{+}\psi_{-}
\nonumber \\
&&+~\bar{\psi_{\dot +}}\bar{\psi_{\dot -}}\left\{\psi_{+},\psi_{-}\right\}
~-~\bar{\psi_{\dot -}}\left\{\bar{\psi_{\dot +}},\psi_{-}\right\}\psi_{+}~.
\end{eqnarray}
Using the anticommutators given in equations (\ref{car}) and the relation
of the incoming and outgoing waves,
\begin{equation}
\psi^{out}=\psi_{+}(u)=\psi_{-}(v'+\delta v_{0})\;\;,\;\;\psi^{in}=\psi_{-}(v)~,
\label{rep}
\end{equation}
the above commutator is
simplified to:
\begin{equation}
2\pi\imath\delta^2(\Omega'-\Omega)
\left[\bar{\psi_{\dot +}}\psi_{-}~-~\bar{\psi_{\dot -}}\psi_{+}\right]
\left[\delta (v'-v)~+~\delta
v_{0}\partial_{v'}\delta (v'-v)\right].
\end{equation}
Which can be further simplified by using equation(\ref{rep}), to
\begin{eqnarray}
~\left[J^{out},J^{in}\right]&=&2\pi\imath \delta (\Omega'-\Omega)\delta v_{0}\left[\partial_{v'}\delta
(v'-v)\left(\bar{\psi_{\dot -}}(v')\psi_{-}(v)~-~\bar{\psi_{\dot -}}\psi_{-}(v)\right)\right .\nonumber\\ 
&&+ \left .\delta (v'-v)\left(\partial_{v'}\bar{\psi_{\dot
-}}(v')\psi_{-}(v)~-~\bar{\psi_{\dot -}}(v)\psi_{-}(v')\right)\right] 
\nonumber \\
&=&
2\pi\imath\delta (\Omega'-\Omega) \delta v_{0}\partial_{v'}
\left[\delta (v'-v)\left(\bar{\psi_{\dot -}}(v')\psi_{-}(v)~-~
\bar{\psi_{\dot -}}(v)\psi_{-}(v')\right)\right] \nonumber \\
&=&0~.
\end{eqnarray}
Hence in the semiclassical approximation the commutator is the
canonical one.

\end{document}